\documentclass[12pt]{article}
\usepackage{latexsym,graphicx,multirow}
\usepackage{amssymb}
\usepackage{amsmath}
\usepackage{amscd}
\usepackage{amsthm}
\usepackage[left=2cm,top=2.5cm,right=2.5cm,bottom=1.5cm]{geometry}
\usepackage{hyperref}
\usepackage{color}

\usepackage{epstopdf}
\usepackage{cite}

\usepackage[utf8]{inputenc}

\begin{document}
    \begin{center}
        \large{\bf{ Kaniadakis Holographic Dark Energy in non-flat Universe}} \\
        \vspace{10mm}
   \normalsize{ Umesh Kumar Sharma$^{a,1}$, Vipin Chandra Dubey$^{b,2}$, A. H. Ziaie$^{c,3}$, H. Moradpour$^{c,4}$}\\
    \vspace{5mm}
    \normalsize{$^a$ Department of Mathematics, Institute of Applied Sciences and Humanities, GLA University
        Mathura-281406, Uttar Pradesh, India}\\
    \vspace{2mm}
    \normalsize{$^b$ Department of Mathematics, 
    	Jaypee Institute of Information Technology, 
    	A-10, Sector-62, Noida-201309,  
    	Uttar Pradesh, India }\\
    \vspace{2mm}
    \normalsize{$^c$ Research Institute for Astronomy and Astrophysics of Maragha (RIAAM), University of Maragheh, P.O. Box 55136-553, Maragheh, Iran}\\
\vspace{2mm}
    $^1$E-mail: sharma.umesh@gla.ac.in\\
    $^2$E-mail: vipindubey476@gmail.com \\
    $^3$E-mail: ah.ziaie@maragheh.ac.ir \\
    $^4$E-mail: hn.moradpour@maragheh.ac.ir \\
    \vspace{10mm}

\end{center}
\begin{abstract}
In present research, we construct Kaniadakis holographic dark energy
(KHDE) model within a non-flat Universe by considering the Friedmann-Robertson-Walker (FRW) metric with open and closed
spatial geometries. We therefore investigate cosmic evolution by employing the density parameter of the dark energy
(DE), the equation of state (EoS) parameter and the deceleration
parameter (DP). The transition from decelerated to accelerated expanding phase for the KHDE Universe is
explained through dynamical behavior of DP. With the classification of matter and
DE dominated epochs, we find that the Universe thermal history can be
defined through the KHDE scenario, and moreover, a phantom regime is
experienceable. The model parameters are constrained by applying
the newest $30$ data cases of $H(z)$ measurements, over the
redshift span $0.07 \leq z \leq 2.36$, and the distance modulus
measurement of the recent Union $2.1$ data set of type Ia supernovae. The
classical stability of KHDE model has also been addressed.\\
\smallskip

\end{abstract}

\section{Introduction}

It was promulgated by Gibbs that the systems involving
interactions (long-range) are no longer extensive
\cite{gibbs1960elementary}, a theory acts as a benchmark of
different entropy formalisms \cite{kaniadakis2001non,
kaniadakis2002statistical,masi2005step,tsallis2013black}. These
entropies provide considerable cosmological models \cite{
Moradpour:2016rcy, Nunes:2015xsa, Komatsu:2016vof,
Moradpour:2017ycq, Abreu:2018pua}. In this regard, new models of
holographic dark energy (HDE)
\cite{tavayef2018tsallis,jahromi2018generalized, Sharma:2020ylh,
moradpour2018thermodynamic, Srivastava:2020cyk} have recently been
proposed using these generalized entropies that have also been
used to study black holes \cite{biro2013q, ghaffari2019black}.
These entropies describe inflation without considering inflaton
\cite{ghaffari2020inflation}, affects the Jeans mass
\cite{Moradpour:2019wpj}, and in general, may be considered as the
sources of modified Newtonian dynamics \cite{Moradpour:2018ima}.
They might also be motivated through the quantum features of
gravity \cite{Moradpour:2019yiq,Barrow:2020tzx}.

An attractive approach for understanding the origin of DE is
holographic dark energy hypothesis, based on the Bekenstein
entropy \cite{Srednicki:1993im} (which takes the kind of degrees
of freedom of horizon \cite{Das:2007sj,pavon2020degrees}), called
ordinary holographic dark energy hypothesis (OHDE)
\cite{li2004model}. By accepting apparent horizon as IR (Infrared)
cut-off, OHDE faces some drawbacks
\cite{li2004model,Myung:2007pn}, whereas apparent horizon is a
suitable decisive boundary from the thermodynamic, and causality
viewpoints
\cite{Hayward:1997jp,Hayward:1998ee,bak2000cosmic,cai2005first,akbar2007thermodynamic,cai2009hawking}.
Contrarily, by taking apparent horizon as IR cut-off, holographic
dark energy (HDE) models based on three generalized entropy, can
provide substantial explanation for the Universe expansion
\cite{tavayef2018tsallis,jahromi2018generalized,
moradpour2018thermodynamic,huang2019stability,universe7030067,ZUBAIR2021153,Bhattacharjee_2021}. Thus, practicing
different entropies, numerous proper models of HDE can be
obtained. Many researchers have examined the compatibility of
these generalized entropies with the zeroth law of thermodynamics
\cite{nauenberg2003critique, moyano2003zeroth, tsallis2004comment,
nauenberg2004reply, li2005different,
abe2006temperature,biro2011zeroth, biro2011publisher}.

Recently, Kaniadakis statistics \cite{
Nunes:2015xsa,abreu2018cosmological} as a generalized entropy
measure \cite{kaniadakis2001non,
kaniadakis2002statistical,masi2005step} has been employed to study
some gravitational and cosmological consequences. The generalized
$\kappa$-entropy (Kaniadakis), a single free parameter entropy of
a black hole is obtained as \cite{moradpour2020generalized}

\begin{eqnarray}
\label{eq8} S_\kappa = \frac{1}{\kappa} \sinh (\kappa S_{BH}),
\end{eqnarray}

\noindent where $\kappa$ is an unknown parameter. Therefore, using
this entropy, and holographic dark energy hypothesis a new model
of DE is introduced as Kaniadakis holographic dark energy (KHDE)
\cite{moradpour2020generalized}, which exposes considerable
properties \cite{moradpour2020generalized, Jawad:2021xsr}.

On the other hand, one of the basic interests in cosmology is the
question concerning the shape or curvature of the Universe, or more precisely
the descriptive geometry of the visible Universe. Though recent observations have provided evidences on the flat geometry, there exist some arguments that support the idea of non-flat geometry due to the contribution of small fraction of positive/negative spatial curvature. One can examine this topic by defining the Universe spatial curvature or a
quantity that describes the variation of the historical geometry
of the Universe from flat space geometry \cite{Vagnozzi:2020dfn}.
Presently, the useful addition mode of spatial curvature to the
Universe energy density is quantified by the curvature parameter $\Omega_K$. In a manner, this ordinary portion is to estimate the
curvature parameter $\Omega_{K}$, while $\Omega_{K}>0$ and $\Omega_{K}<0$ contributes to a spatially open and closed
Universe, respectively. Besides holding these ideas in mind, polarization anisotropy data
and Planck 2018 CMB temperature singly introduced the constraint
$\Omega_{K}=-0.044^{+0.018}_{-0.015}$ \cite{Aghanim:2018eyx}, here
$\Omega_{K}=0$ corresponds to a spatially flat Universe. Moreover,
estimated from the BOSS DR12 CMASS example of linking P18 data
with the full-shape (FS) universe power spectrum and including
$\Omega_{K} = 0.0023 \pm 0.0028$ received from this alliance
\cite{Vagnozzi:2020zrh}, one can achieve less constraining
outcomes. It is also notable that in non-flat geometry, the closed Universe models exhibit substantially higher lensing amplitudes in comparison to $\Lambda$CDM model~\cite{Aghanim:2018eyx}. In either situation, from modern and
future cosmic views, the influence of spatial curvature on the evolution of the Universe is the cause why a large number of researches have been attracted to find possible constraints on $\Omega_{K}$~\cite{DiValentino:2020hov}.

\par
Keeping these studies in mind, our aim in the present study is to construct a KHDE model in a non-flat
universe. The work is organized as follows: In Section \ref{S2} the fundamental
equations of KHDE in a non-flat Friedmann-Lema$\hat i$tre-
Robertson- Walker (FLRW)  Universe are presented. In section \ref{S3}, the specific
research on cosmic nature is continued, and we mainly focus on the
EoS and DE density parameters. Section \ref{S4} exposes the techniques
used for data analysis in this study. We examine the stability of
the model in Section \ref{S5}. Lastly, we recap our outcomes in Section
\ref{S6}.

\section{KHDE in Non-Flat Universe}\label{S2}

The cosmic principles and General Relativity (GR) are our basic hypotheses to explain the Universe on large scales. The line element for a homogeneous and isotropic non-flat Universe is presented through the FLRW metric, given by
\begin{eqnarray}
    \label{eq12}
    ds^{2} = -dt^{2}+a^{2}(t)\Big(\frac{dr^{2}}{1-Kr^{2}}+ + r^{2}d\Omega^{2}\Big),
\end{eqnarray}
where $a(t)$ is the scale factor which expresses the increase/decrease of the isotropic and homogenous spatial parts. The constant parameter $K$ represents the spatial curvature with $K=-1,0,1$ respectively corresponds to open, flat and closed Universe models. The spatial curvature efficiently appears as an extra matter-energy term within the Friedmann equations with sectional contribution quantified through the curvature parameter $\Omega_{K} \equiv -\frac{K}{(H_{0}a_{0})^{2}}$, where the Hubble parameter $H$ describes the Universe expansion rate and $0$ indicates the size estimated presently. In general the estimated Hubble parameter $H_{0}$ is introduced as the Hubble constant. Particularly, $K$ and $\Omega_{K}$ appear by opposing signs, hence $\Omega_{K}<0$ represents a closed Universe and $\Omega_{K}>0$ an open one.\\

The HDE states that if DE is supposed to control the present accelerated expansion of the Universe, then, considering the Kaniadakis black hole entropy Eq. (\ref{eq8}),  the amount of vacuum energy accumulated within a box of size $L^3$ must not exceed the energy of its same size black hole \cite{li2004model}. One then gets
\begin{equation}
    \label{eq9}
\Lambda^{4}\equiv   \rho _D\propto \frac{S_{\kappa}}{L^4},
\end{equation}
for the vacuum energy $\rho _D$. Now, as a general limit on the system under study the apparent horizon is considered as the IR cutoff for the FLRW Universe. It then can be located as ~\cite{sheykhi2018modified,hayward1999dynamic,hayward1998unified,bak2000cosmic}

\begin{eqnarray}
    \label{eq10}
    \tilde{r}_A=\frac{1}{\sqrt{\frac{K}{a^2}+H^2}},
\end{eqnarray}
whence the vacuum energy reads
\begin{equation}
\label{eq11}
\rho _D=\frac{3 C^2 H^4}{8\pi\kappa}\left(\Omega _K+1\right)^2 \sinh \left(\frac{\pi  \kappa }{H^2 \left(\Omega _K+1\right)}\right),
\end{equation}
where we have set $L = \tilde{r}_A$ and $C^2$ is an undefined standard constant~\cite{li2004model}. Clearly, for the case $\Omega_{K}=0$, Eq. (\ref{eq11}) coincides with KHDE in a flat Universe~\cite{moradpour2020generalized}. Also, this is
apparent that whenever $\kappa\rightarrow0$ and $\Omega_{K}=0$ , we have $\rho_{D}= \frac{3C^{2}H^{2}}{8\pi}$,
the famous Bekenstein entropy-based HDE (i.e. OHDE)
\cite{li2004model}. In a non-flat Universe, the first and second Friedmann equations including KHDE and dark matter (DM), are given by
\begin{eqnarray}
\label{eq13}
H^2 + \frac{1}{a^2} K=\frac{1}{\tilde{r}^2_A}=\frac{8 \pi }{3} G \left(\rho_D+\rho _m\right),
\end{eqnarray}

\begin{eqnarray}
\label{eq13a}
H^{2} + \frac{2}{3}\dot H + \frac{K}{3a^2}=-\frac{8 \pi }{3} G p_{D},
\end{eqnarray}
where the energy density of KHDE, its pressure and the matter energy density are presented as $\rho_{D}$, $p_{D}$ and $\rho_{m}$, respectively. Working on the fractional densities, the corresponding density parameters for pressure-less fluid, KHDE and spatial curvature are defined as
\begin{eqnarray}
    \label{eq14}
     \Omega_{m} = \frac{8\pi G}{3H^{2}} \rho_{m}, \hspace{1cm} \Omega_{D} = \frac{8\pi G}{3H^{2}} \rho_{D},  \hspace{1cm} \Omega_{K} =\frac{1}{a^{2}} \frac{K}{H^{2}},
\end{eqnarray}
by the virtue of which Eq. (\ref{eq13}) can be recast into the following form
\begin{eqnarray}
\label{eq15}
\Omega _K +1 = \Omega _m +\Omega _D.
\end{eqnarray}
The conservation law for KHDE and matter leads individually to the following equations

\begin{eqnarray}
\label{eq17}
\dot \rho_{D} +3 (1+\omega_D)\rho_{D} H = 0,
\end{eqnarray}
\begin{eqnarray}
\label{eq16}
\dot \rho_{m} + 3  \rho_{m} H = 0,
\end{eqnarray}
where $H =\frac{\dot a}{a}$ and KHDE EoS prameter is shown as $
\omega _D = p _D/\rho _D$. In order to obtain the behavior of DP we proceed with differentiating Eq. (\ref{eq11})
with respect to cosmic time $t$. One then finds
\begin{eqnarray}
\label{eq18}
{\dot\rho _D}= 2 H^3 \left(\frac{\dot{H}}{H^2}-\Omega _K\right) \left(\frac{2 \rho _D}{H^2 \left(\Omega _K+1\right)}-\frac{3}{8} C^2 \cosh \left(\frac{\pi  \kappa }{H^2 \left(\Omega _K+1\right)}\right)\right).
\end{eqnarray}
Now, taking the derivative of Eq. (\ref{eq13}) along with using Eqs. (\ref{eq15})-(\ref{eq18}) gives
\begin{eqnarray}
\label{eq19}
\frac{\dot{H}}{H^2}= \frac{-3}{2}\left(\omega _D \Omega _D+\frac{\Omega _K}{3}+1\right),
\end{eqnarray}
whereby one finds the following expression for DP
\begin{eqnarray}
\label{eq20}
q =  -\frac{\dot{H}}{H^2}-1 =\frac{1}{2} \left(3 \omega _D \Omega _D+\Omega _K+1\right).
\end{eqnarray}
We note that this result can be also obtained from Eq. (\ref{eq13a}). This is due to the fact that Eqs. (\ref{eq13}) and (\ref{eq13a}) along with conservation equations are not independent from each other. We can also get the behavior of EoS parameter by solving Eq. (\ref{eq17}) with respect to $\omega_{D}$ and substituting for $\rho_{D}$ and $\dot{\rho} _D$ from Eqs. (\ref{eq11}) and (\ref{eq18}), respectively. By doing so, one arrives at the following expression for EoS parameter

\begin{eqnarray}
\label{eq21}
\omega _D= \frac{1}{3} \left(\frac{2 \pi  \left(H^2+\dot{H}\right) \kappa  \coth \left(\frac{\pi  \kappa }{H^2 \left(\Omega _K+1\right)}\right)}{H^4 \left(\Omega _K+1\right){}^2}-\frac{2 \left(\pi  \kappa  \coth \left(\frac{\pi  \kappa }{H^2 \left(\Omega _K+1\right)}\right)+2 \left(H^2+\dot{H}\right)\right)}{H^2 \left(\Omega _K+1\right)}+1\right).
\end{eqnarray}
Additionally, from the second part of Eq. (\ref{eq14}) and Eq. (\ref{eq11}) the density parameter of KHDE can be expressed as
\begin{eqnarray}
    \label{eq22}
    \Omega _{D}=\frac{C^2}{\kappa} H^2 \left(\Omega _K+1\right){}^2 \sinh \left(\frac{\pi  \kappa }{H^2 \left(\Omega _K+1\right)}\right).
\end{eqnarray}
As expected for the flat case $\Omega_{K}=0$, Eq. (\ref{eq22}) reduces to the expression found in~\cite{moradpour2020generalized}.

\section{Cosmological evolution}\label{S3}
 In this section, we continue to study the cosmic evolution of KHDE for both open and closed Universe models.
As we mentioned above, Eq. (\ref{eq22}) determines the behavior of the density parameter of DE for the two spatially curved cases. One can easily express the cosmic evolution in terms of the more convenient redshift parameter, setting the current scale factor value to $a_{0}=1$. We elaborate equations numerically, imposing the initial conditions $\Omega _D = 0.70$ and  $H _0=67.9$ in agreement with recent observations \cite{Aghanim:2018eyx}.\\
\begin{figure}
    \begin{center}
        \includegraphics[width=8cm,height=6cm, angle=0]{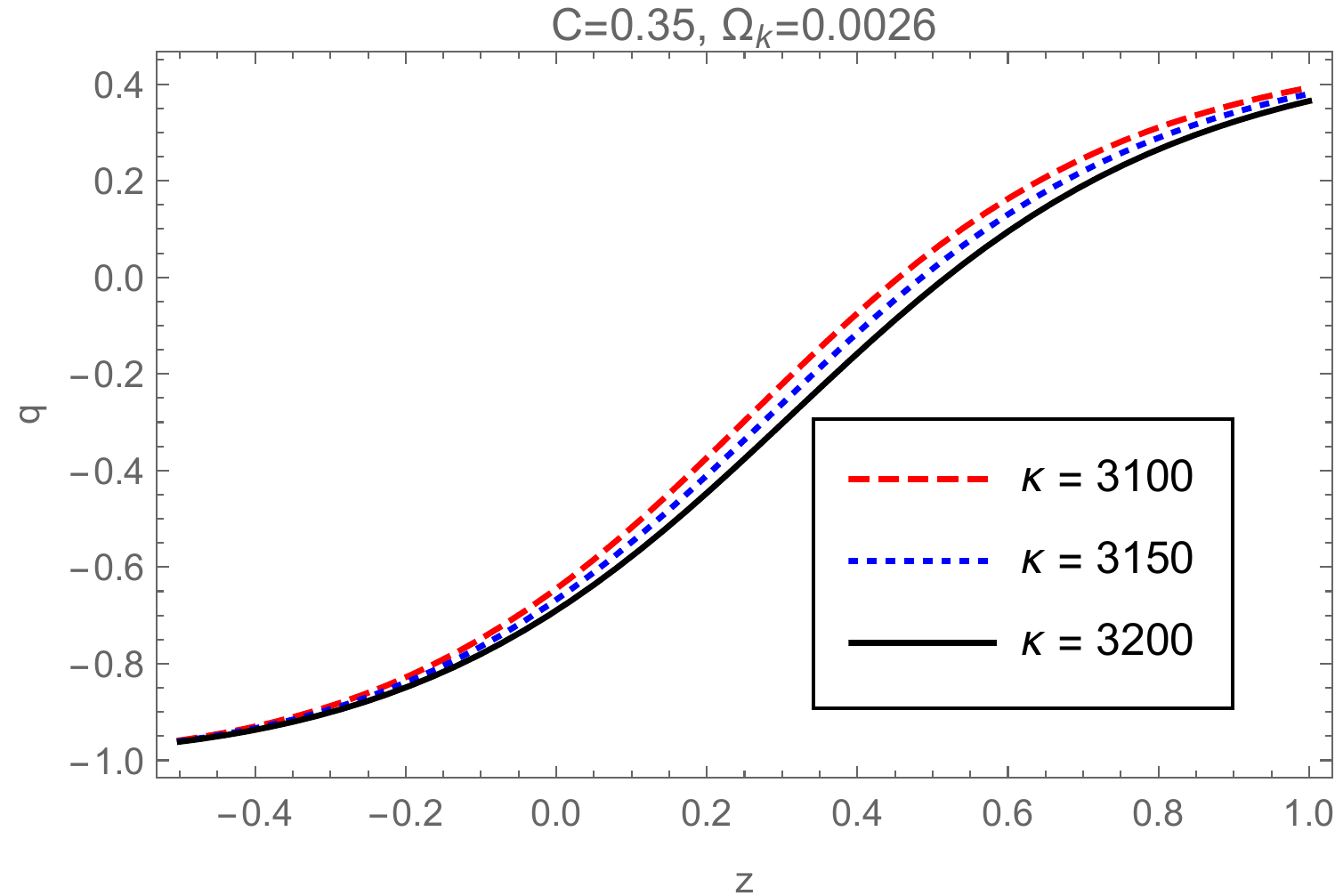}
        \includegraphics[width=8cm,height=6cm, angle=0]{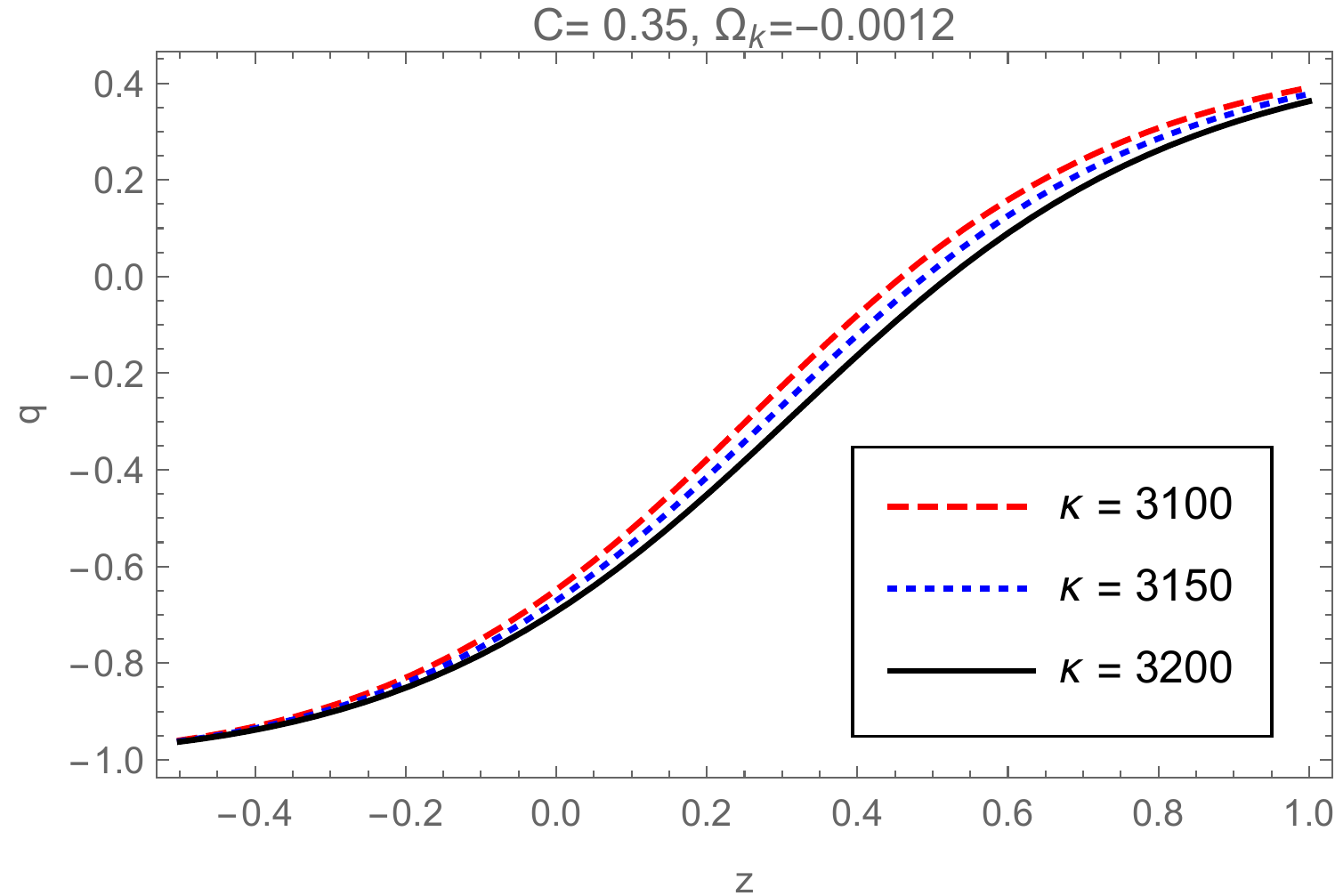}
        \caption { The evolution trajectories of deceleration parameter $q$ verses redshift $z$ for $C= 0.35$ and different values of model parameter $\kappa$ with initial conditions $\Omega_{D0}$= $0.70$ and $H_{0}=67.9$. The left panel has been plotted for an open Universe with $\Omega_{K}$= $0.0026$ and the right one has been plotted for a closed Universe with $\Omega_{K}$= $ -0.0012$.} \label{FIG1}
    \end{center}
\end{figure}
\begin{figure}
    \begin{center}
        \includegraphics[width=8cm,height=6cm]{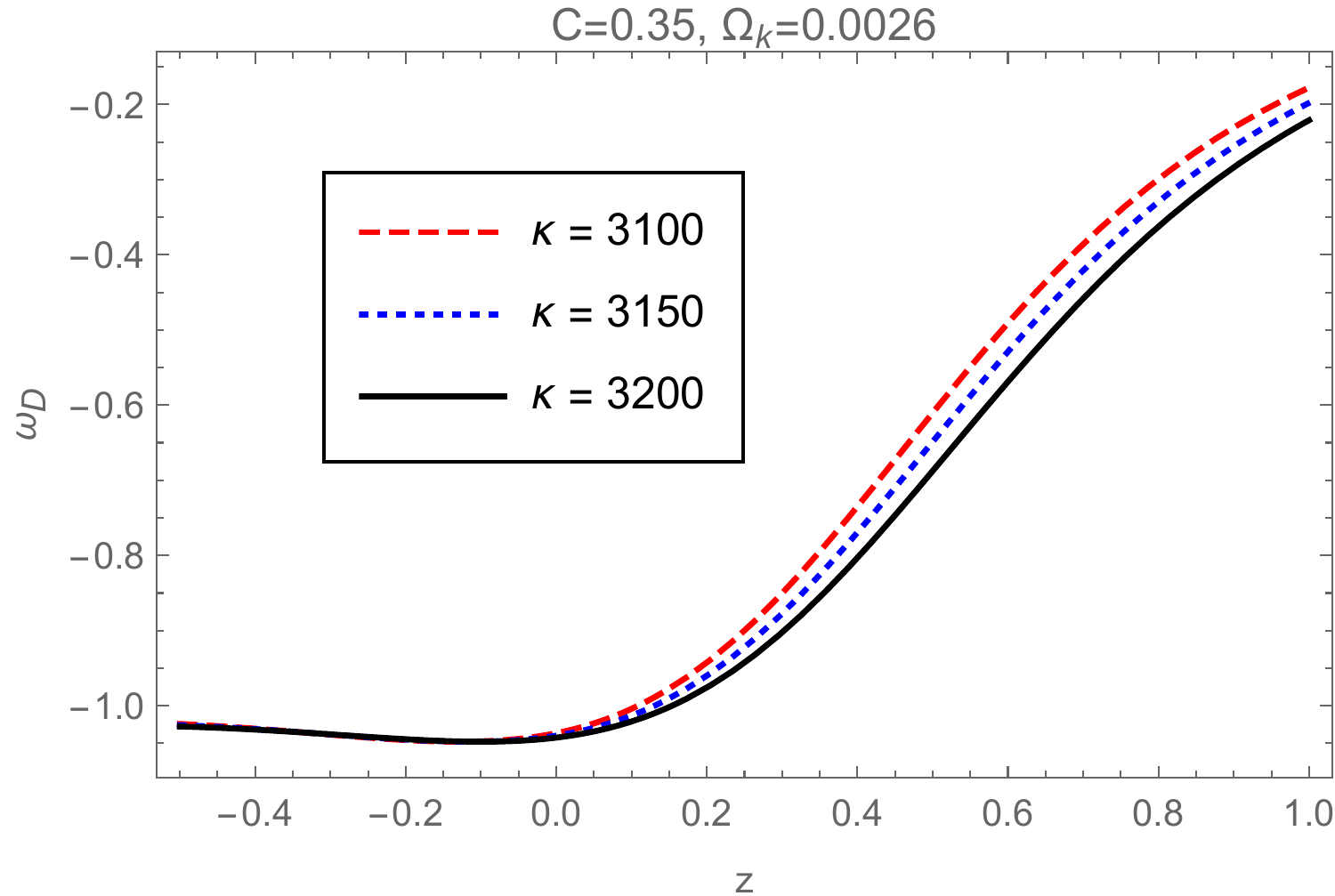}
        \includegraphics[width=8cm,height=6cm]{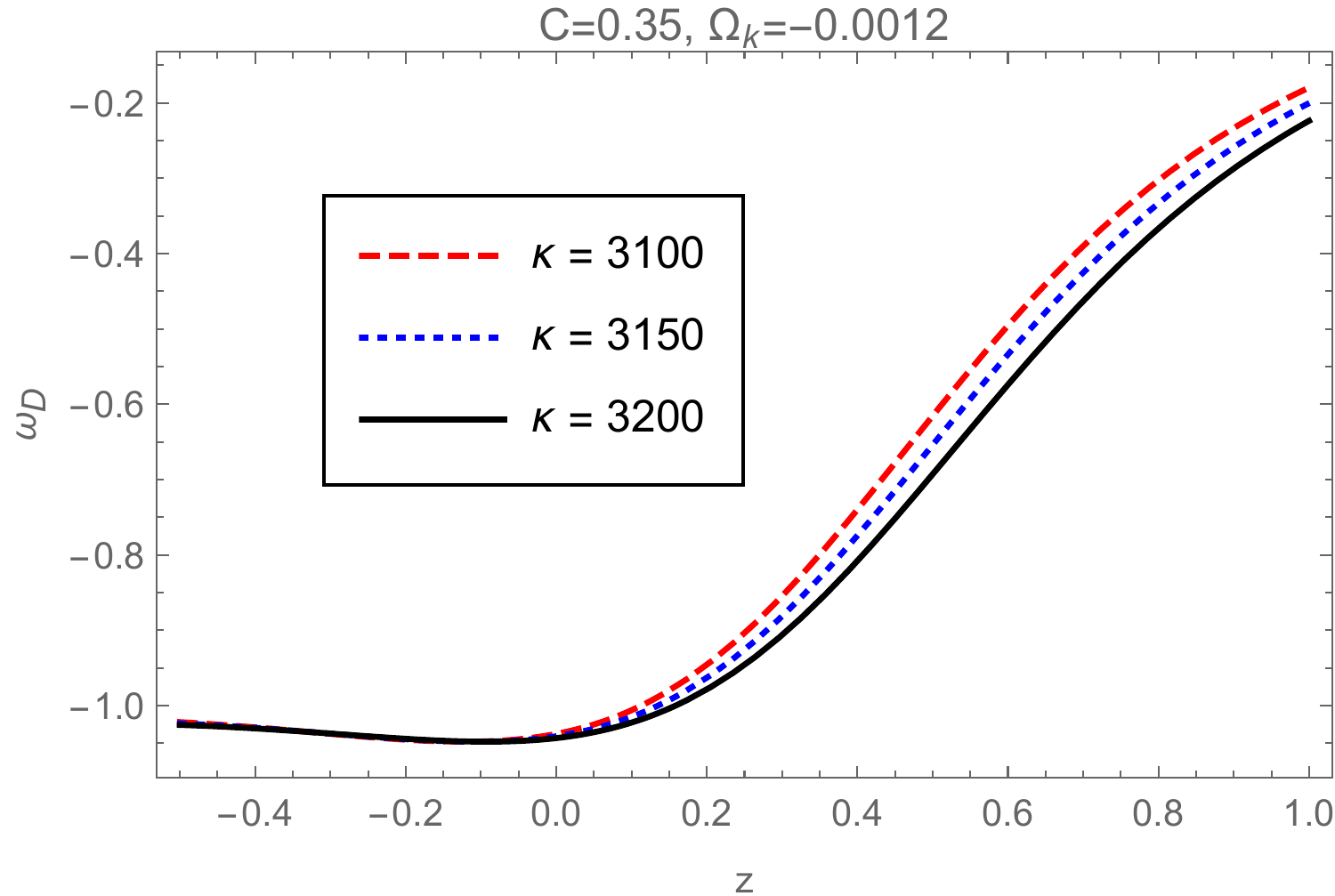}
        \caption{Behavior of EoS parameter $\omega_D$ versus redshift $z$ for $C= 0.35$ and different values of model parameter $\kappa$ with initial conditions $\Omega_{D0}$= $0.70$ and $H_{0}=67.9$ for open Universe $\Omega_{K}$= $0.0026$ (left panel) and closed Universe $\Omega_{K}$= $ -0.0012$ (right panel).}\label{FIG2}
    \end{center}
\end{figure}
\begin{figure}
    \begin{center}
        \includegraphics[width=8cm,height=6cm]{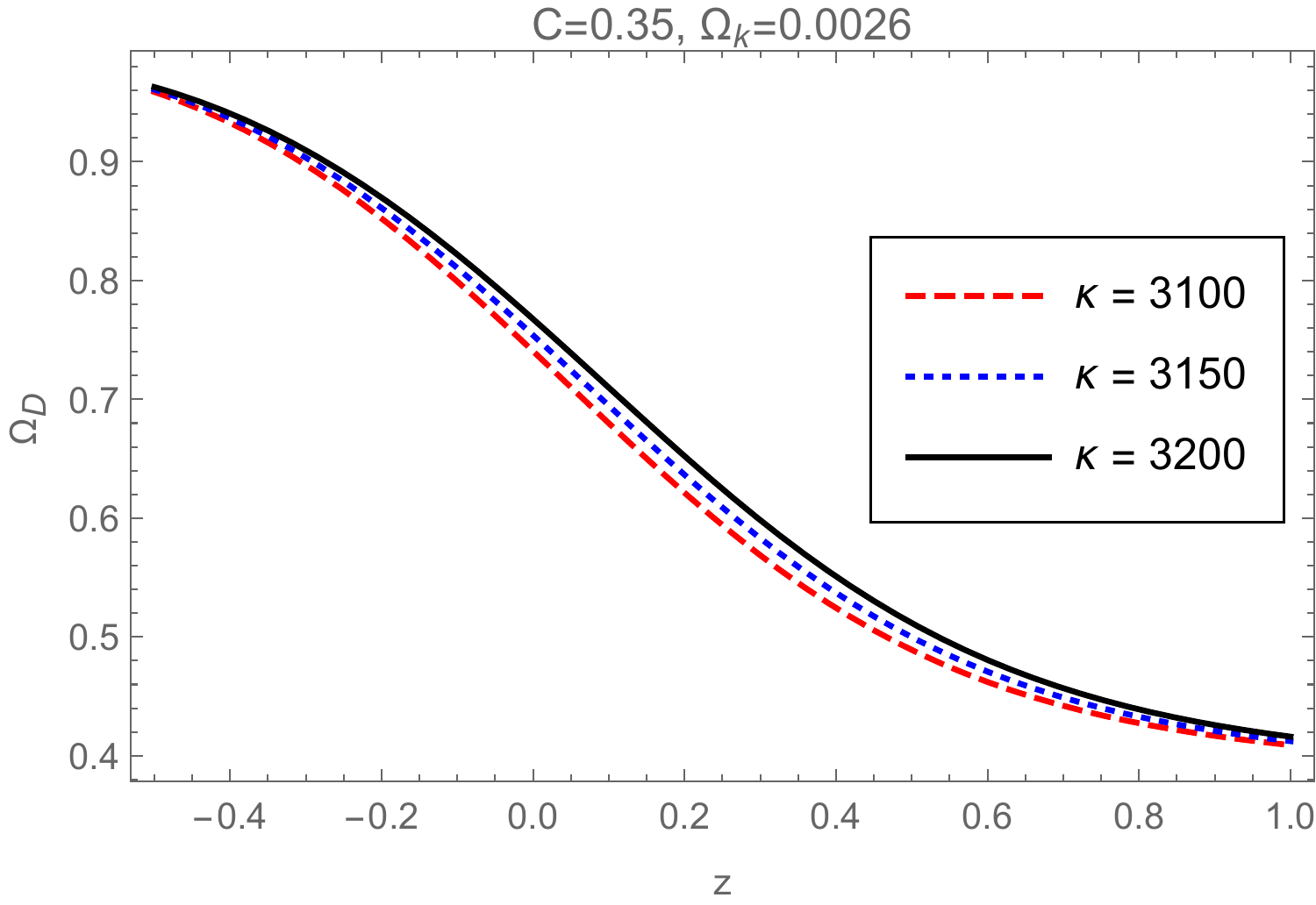}
        \includegraphics[width=8cm,height=6cm]{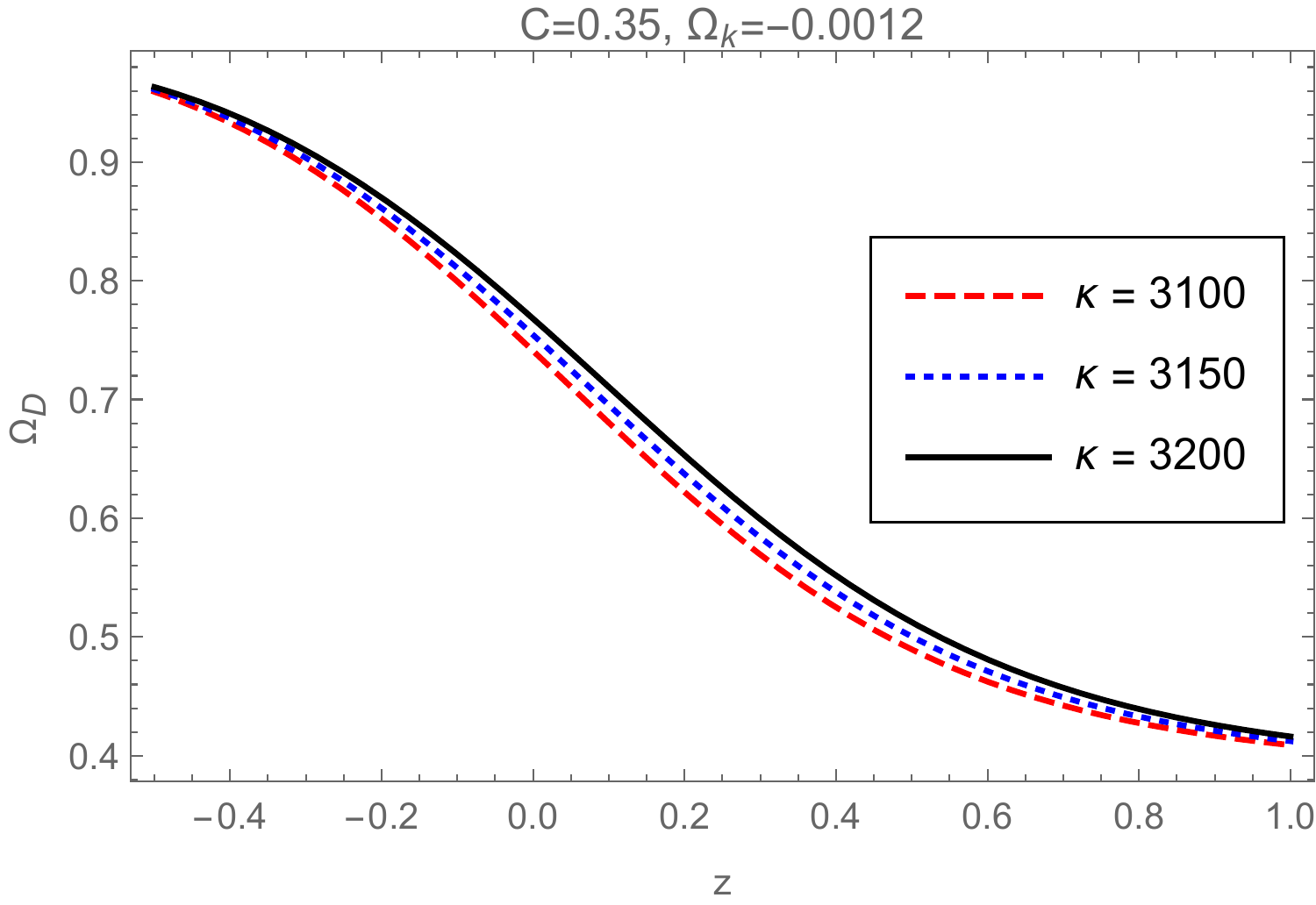}
        \caption{ Behavior of density parameter $\Omega _D$ versus redshift $z$ for an open (left panel) and a closed (right panel) Universe for $C= 0.35$ and different values of model parameter $\kappa$ with initial conditions $\Omega_{D0}$= $0.70$ and $H_{0}=67.9$.}\label{FIG3}
    \end{center}
\end{figure}
The behavior of deceleration parameter $q$ for as a function of redshift $z$, can be obtained by examining Eq. (\ref{eq20}) which is solved numerically by taking the initial conditions stated above. The numerical solution of $q$ is depicted in Fig. (\ref{FIG1}). We have taken different observationally best-fitted values of KHDE model parameter $\kappa = 3100,\,3150$ and $3200$ and $C= 0.35$ for an open Universe $\Omega_{K0}$= $0.0026$ (left panel) and close Universe $\Omega_{K0}$= $ -0.012$ (right panel). We observe that the evolution of both closed as well as open Universe models from early decelerating to late-time accelerating phase of expansion occurs at transition redshift $z_{tr}$ where $q(z_{tr})=0$. The transition redshift for the present model is obtained as, $z_{tr}\approx0.5$ as required from observations.

In order to study the nature of the EoS parameter of KHDE and specifically to investigate whether it is influenced by $\kappa$ and $\Omega_{K}$ parameters, we plot $\omega_{D}$ against redshift in Fig. (\ref{FIG2}) for different values of $\kappa$ parameter and for open (left panel) and closed (right panel) Universe models. As we can see, for different values of $\kappa$ parameter, the evolution of $\omega_{D}$ and its prevailing value $\omega_{D}(z = 0) \equiv \omega_{D0}$ tend to assume more moderate values. Interestingly, for $z>0$ the EoS parameter of DE lies completely within the quintessence region. As the Universe evolves to redshift $z\leq0$, the phantom divide ($w_{DE}=-1$) is crossed at a certain value of redshift. This value of the redshift depends on the model parameters of KHDE model. Therefore, in the framework of present model, it is possible for the EoS of KHDE to cross to the phantom regime, contrary to the case of standard HDE. \\

In Fig. (\ref{FIG3}) we depict the behavior of DE density parameter $\Omega_{D}$ during the evolution of an open (left panel) and a closed (right panel) Universe for different values of the Kaniadakis parameter $\kappa$. Since we witnessed the general thermal history of the Universe through the classification of matter and radiation dominated eras, the transition from matter to DE domination complies with the required scenario of structure formation of the Universe. It is remarkable that for more transparency we should extend the evolution of the Universe to the distant futurity, from where we can observe the outcomes of cosmic evolution in complete DE domination as expected. It is also noteworthy that in comparison to the flat KHDE model, where the phantom regime was not obtained, the incorporation of spatial curvature can drive the Universe into the phantom region, which can be an interesting subject of research and discussion in the future.
\begin{figure}[htp]
    \begin{center}
        \includegraphics[width=8cm,height=6cm]{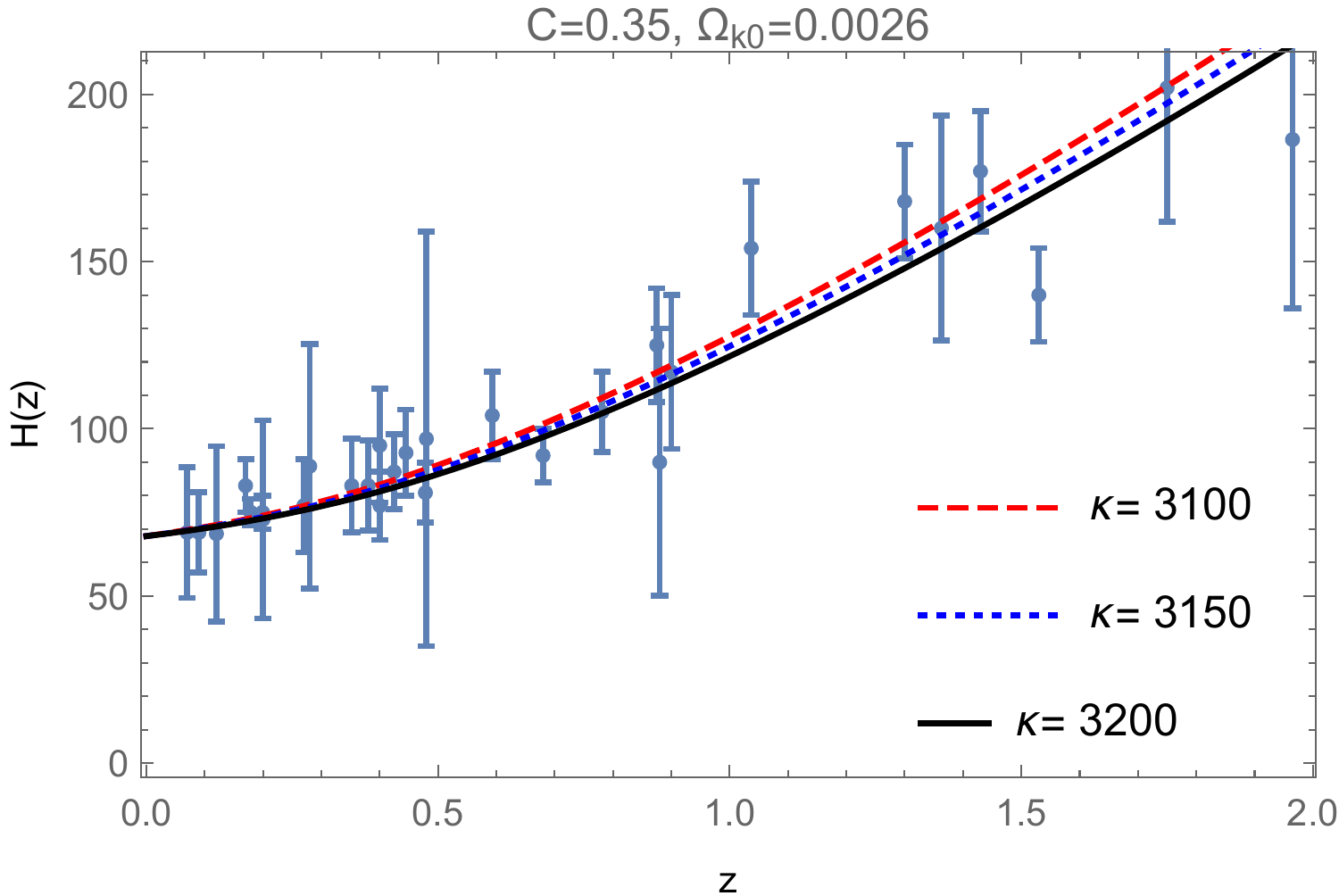}
        \includegraphics[width=8cm,height=6cm]{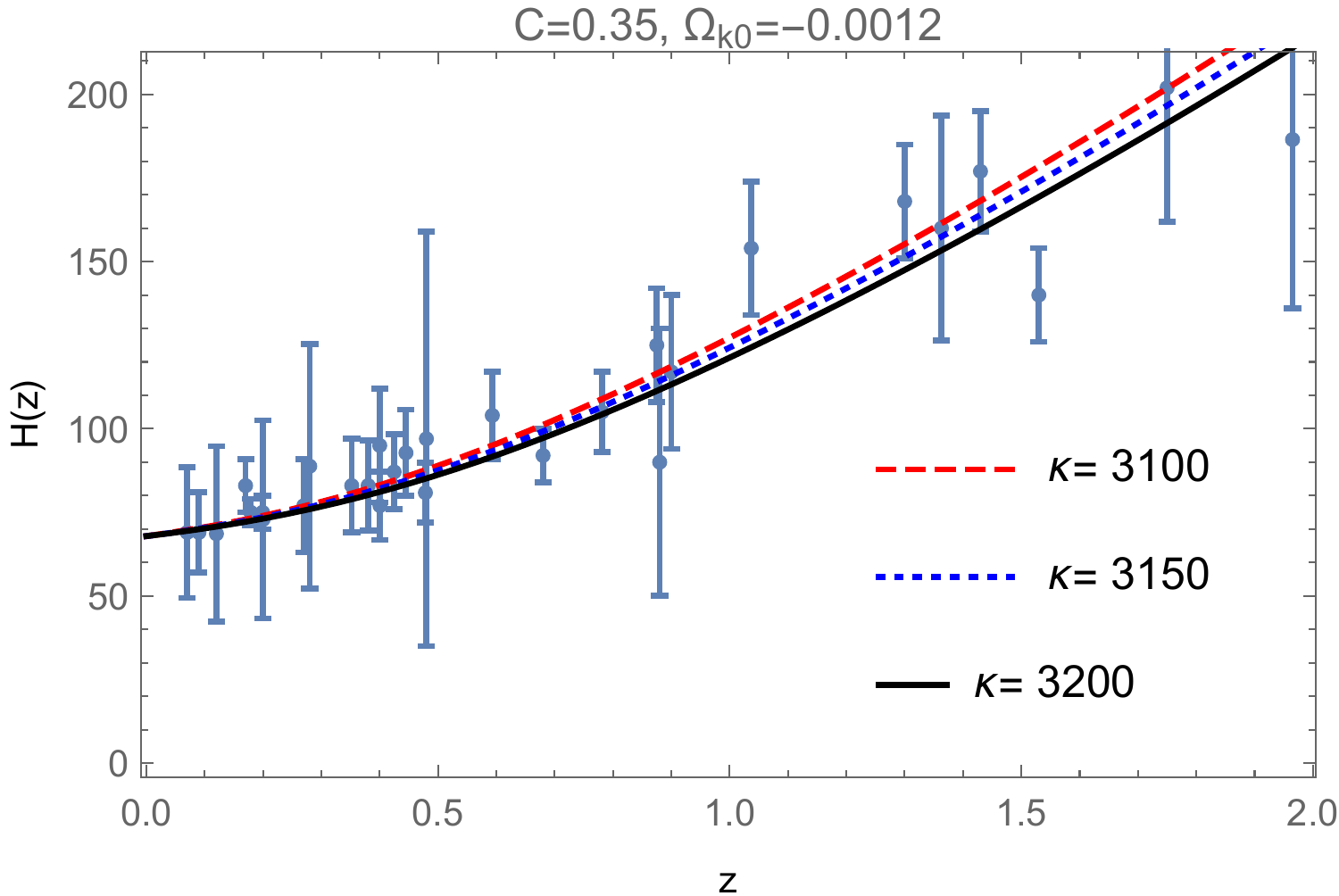}
        \caption{The evolution of $H(z)$  versus redshift $z$ with error bars for $C= 0.35$ and different values of model parameter $\kappa$ with initial conditions $\Omega_{D0}$= $0.70$ and $H_{0}=67.9$ for open Universe $\Omega_{K}$= $0.0026$ (left panel) and  closed Universe $\Omega_{K}$= $-0.0012$ (right panel). In this plot, the dots correspond to the 30 $H(z)$ data points}.\label{FIG4}
    \end{center}
\end{figure}

\section{ Observational data analysis }\label{S4}

This section deals with the most advanced cosmological investigations of KHDE model. We have explained the datasets considered in our analysis in-brief, specifically, the current data of Hubble parameter has been achieved through the CC (cosmic chronometer) mode and Type Ia Supernova (SNIa).

\subsection{Observational Hubble Data (OHD)}

We take Hubble datasets $H(z)$ from table 4 of Ref. \cite{moresco20166}, which contains 30 $H(z)$ observational dataset limits spanning the redshift interval of $0.07 \leq z \leq 1.965$. These uncorrelated data can be obtained by the cosmic chronometric (CC) technique. The purpose of taking these set of data lies in the case that the OHD dataset acquired of the CC method is individualistic. The CC data of passively evolving Universes depends on the method of distinct dating of galaxies. The evolution of $H(z)$ (Hubble parameter) for our KHDE model has been shown in Fig. (\ref{FIG4}), for both closed and open Universes, and compared with the newest 30 points of $H(z)$ data~\cite{moresco20166}. We observe that the present KHDE model is fully compatible with the OHD dataset at variance with the redshift parameter.

\subsection{Distance modulus  $\mu$}

To examine the cosmological models, the datasets extracted from the SNIa are very beneficial particularly as a primary evidence for accelerated Universe. Therefore, we have further used the distance modulus dataset sample of 580 scores of Union 2.1 combined with SNIa~\cite{suzuki2012hubble}, besides the CC data. To investigate the evolution of the Universe, there is a prominent observational tool namely the redshift-luminosity distance relation \cite{liddle2000cosmological}. Due to the expansion of the universe, the light reaching out towards a distant luminescent body becomes redshifted and we can get the equation for the luminosity distance (DL) in terms of $z$. With help of luminosity distance, we can obtain the flux of a source, which is presented as 
\begin{eqnarray}
    \label{eq23}
    D_L = a_0r(z + 1),
\end{eqnarray}
here $r$ denotes the radial coordinate of the source. Copeland et al.\cite{copeland2006m} have formulated DL as given below 
\begin{eqnarray}
    \label{eq24}
    D_L=
    \frac{c(1 + z)}{H_0}\int_{0}^{z}\frac{dz}{h(z)}, \quad
    h(z) = \frac{H}{H_0}.
\end{eqnarray}
The distance modulus $\mu$ is then obtained as~\cite{copeland2006m} 
\begin{eqnarray}
\label{eq25}
\mu  = 25 +  5\log_{10}\left( \frac{D_L}{M_{pc}}\right).
\end{eqnarray}
From Eq. (\ref{eq24}), one gets the following expression for distance modulus
\begin{eqnarray}
\label{eq26}
\mu = 25 +  5\log_{10}\left[\frac{c(1 + z)}{H_0}\int_{0}^{z}\frac{dz}{h(z)}\right].
\end{eqnarray}

\begin{figure}

\includegraphics[width=8cm,height=6cm]{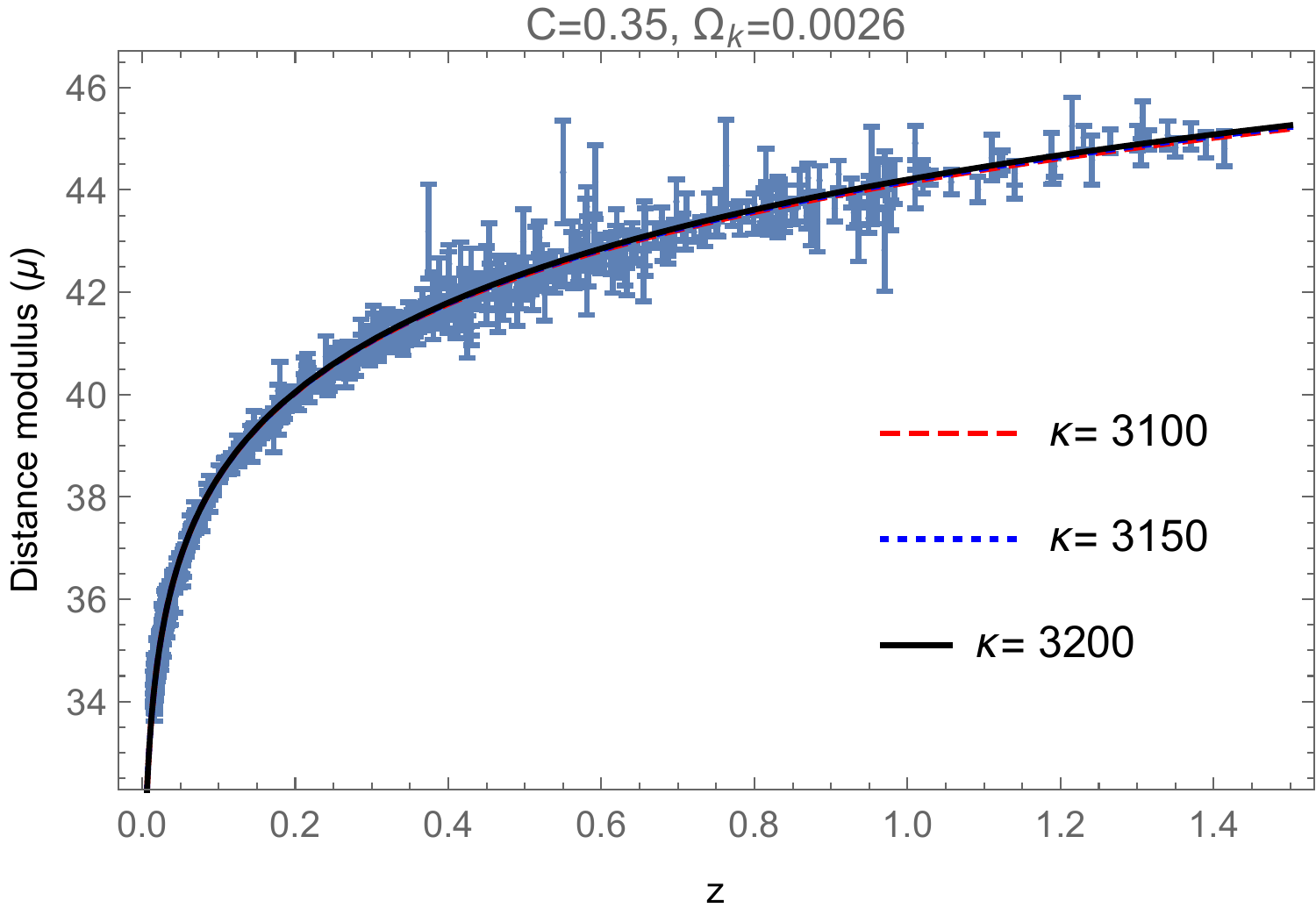}
\includegraphics[width=8cm,height=6cm]{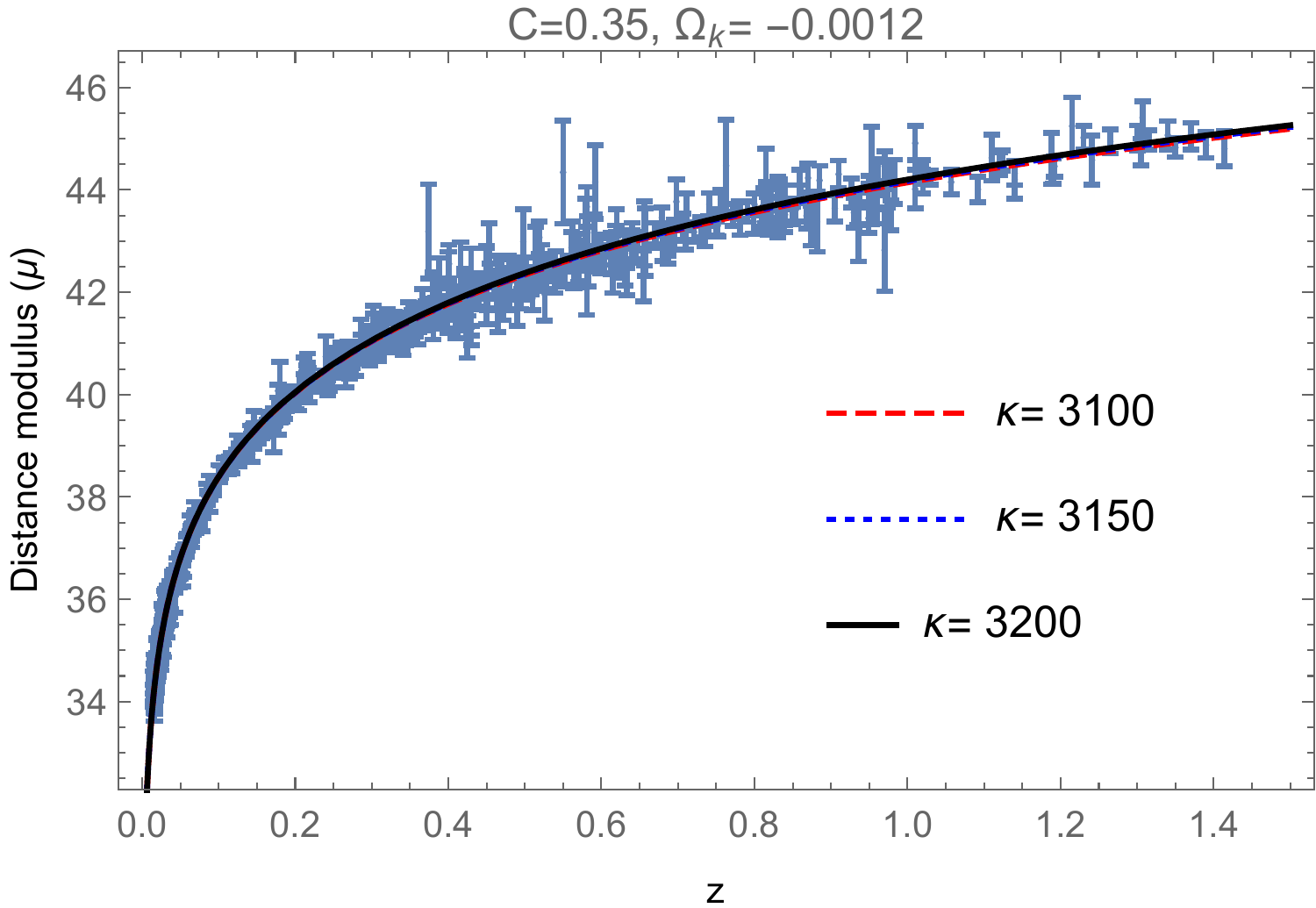}
    \caption {Evolution of Luminosity Distance versus redshift $z$ for $C= 0.35$ and different values of model parameter $\kappa$ with initial conditions $\Omega_{D0}$= $0.70$ and $H_{0}=67.9$ for open Universe $\Omega_{K}$= $0.0026$ (left panel) and closed Universe $\Omega_{K}$=$-0.0012$ (right panel).}\label{FIG5}

\end{figure}

    \begin{figure}[htp]
    \begin{center}
        \includegraphics[width=8cm,height=6cm]{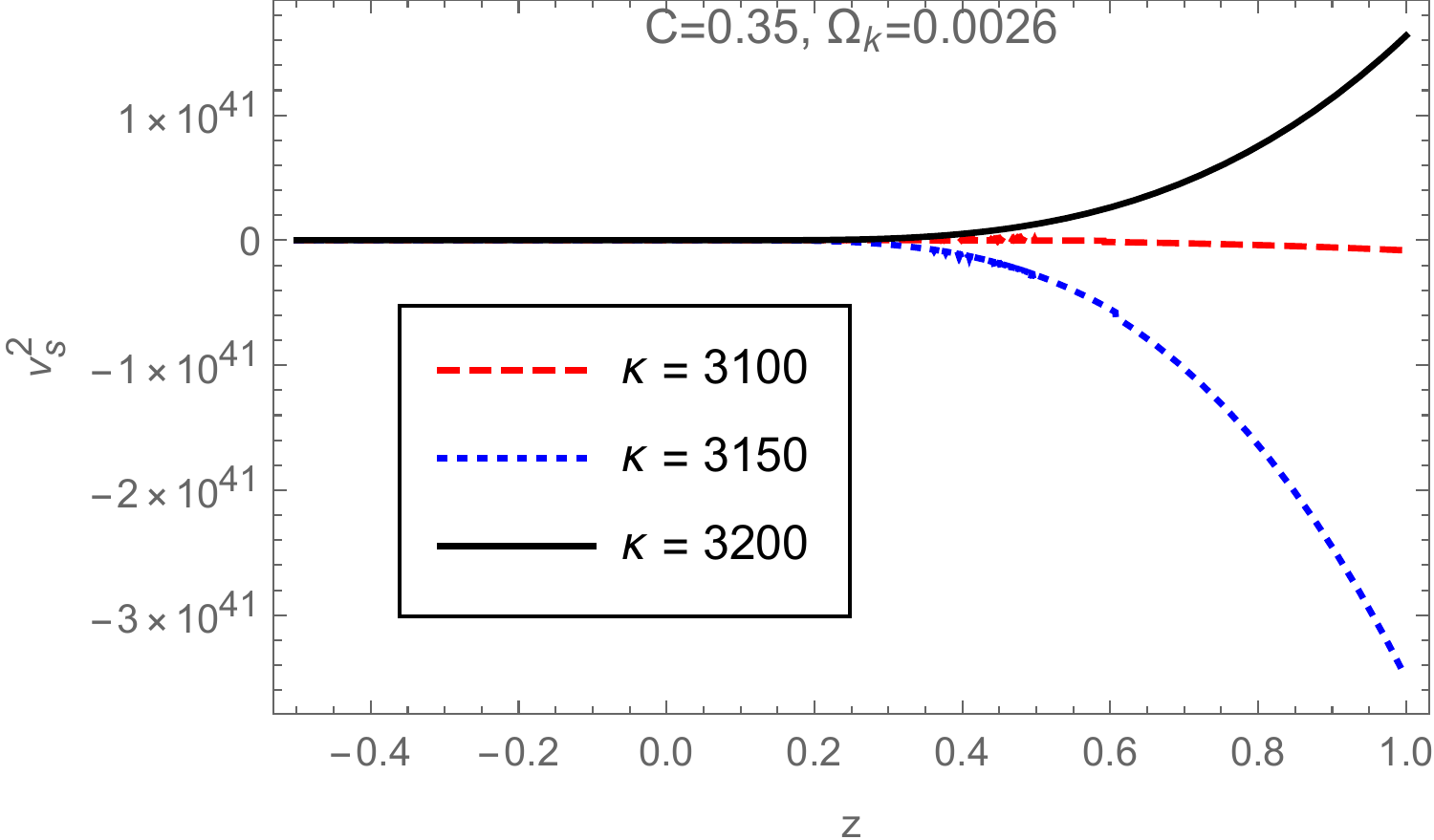}
        \includegraphics[width=8cm,height=6cm]{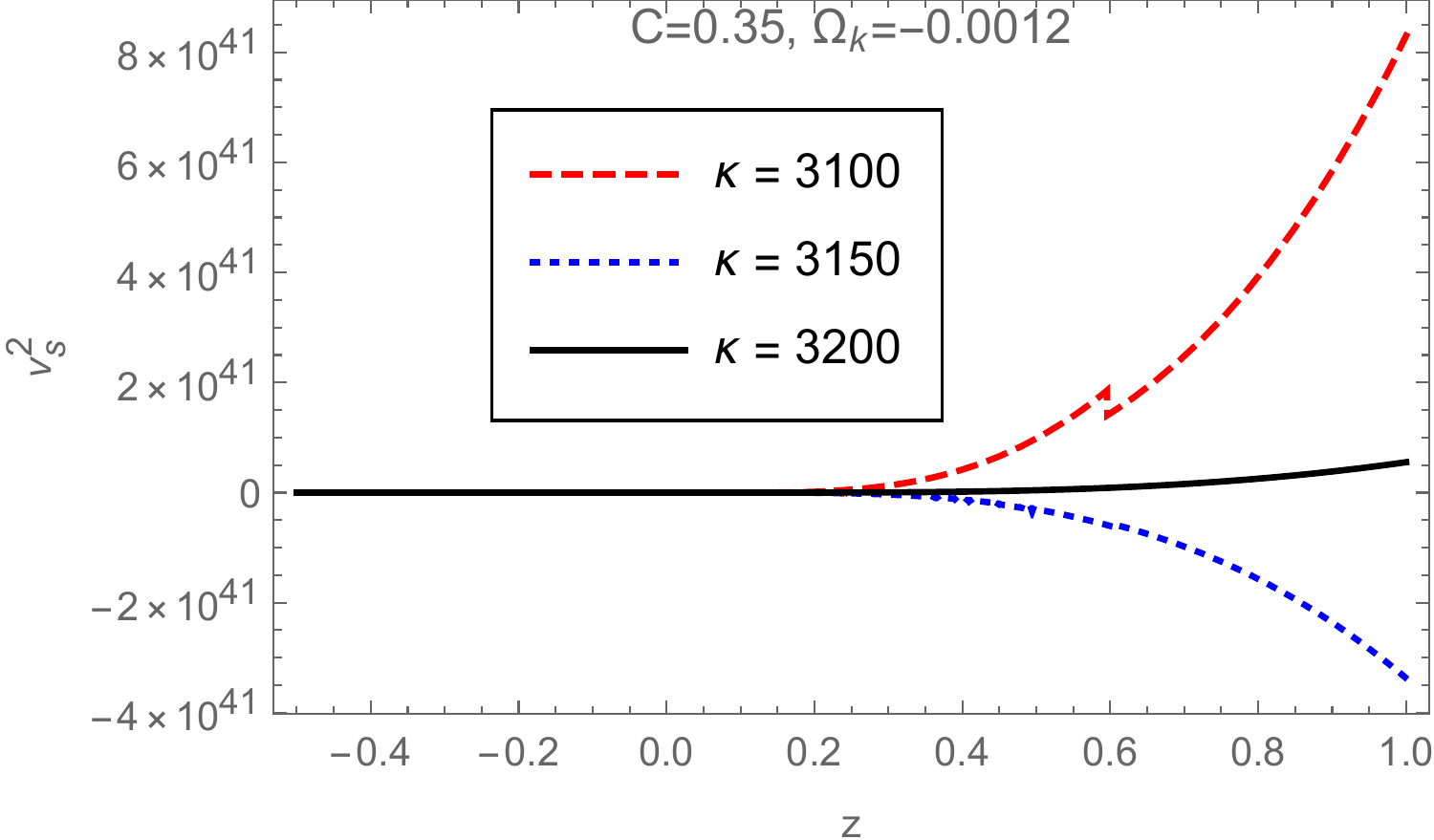}
        \caption{The behavior of $v_{s}^{2}$ against redshift $z$ for $C= 0.35$ and different values of model parameter $\kappa$ with initial conditions $\Omega_{D0}$= $0.70$ and $H_{0}=67.9$ for open Universe $\Omega_{K}$= $0.0026$ (left panel) and closed Universe $\Omega_{K}$= $-0.0012$ (right panel).}
        \label{FIG6}
    \end{center}
\end{figure}
Figure (\ref{FIG5}) shows the most suitable period of distance modulus $\mu$(z) in terms of $z$ for KHDE model combining the 580 scores of Union 2.1 \cite{suzuki2012hubble} with SNIa data. This figure displays the error graph for KHDE model presented in solid line by applying 580 scores of Union 2.1 combined with SNIa datasets. Also from this figure one could easily witness the moderate reflection of KHDE model on the observed values of distance modulus $\mu$ for every data point, which indeed justifies our model.
\section{Stability}\label{S5}
The present section deals with examining classical stability of KHDE model through the squared speed of sound ($v_{s}^{2}$) for a closed as well as open FRW Universe. The squared speed of sound is given as ~\cite{calabrese2011future,vagnozzi2020we}
    \begin{eqnarray}
        \label{eq31} v_s^2=\frac{dp_{D}}{d \rho_{D}}= \frac{\rho _D }{\dot{\rho}_D}\dot\omega _D +\omega _D.
    \end{eqnarray}
In figure (\ref{FIG6}), we plotted the behavior of squared speed of sound ($v_{s}^{2}$) against redshift using different values of $\kappa$ parameter. As we see for some values of Kaniadakis parameter namely, $\kappa=3150$, we have $v_{s}^{2}<0 $ and thus, KHDE model is not stable in both open and closed Universe models, see the blue dotted curve. However, for other values of $\kappa$ parameter $v_{s}^{2}>0$ and the KHDE model is stable for late-time, see the black solid and red dashed curves. Hence the condition on classical stability of the model puts limit on the allowed values of Kaniadakis parameter.

\section{Conclusions}\label{S6}

In this research, we have examined KHDE model in the non-flat universe which is dependent on two model parameters $C$ and $\kappa$. While estimating these model parameters we have also used the dataset samples of distance modulus of 580 scores of Union 2.1 combining SNIa along with the $30$ CC data points of $H(z)$ analyses. Besides, we have reconstructed the evolving behavior of $q(z)$, $\omega_{D}$ and $\Omega_{D}$ for the existing model, while working on the fittest values of $C$ and $\kappa$.\\

Proceeding to the detailed investigation, we showed that the classification of DE and matter dominated eras through the supporting diagrams can explain the Universe's thermal history. Furthermore, we examined the effect of the Kanadakis parameter $\kappa$, as well as of the curvature density parameter at present, on the EoS parameter of DE. As we observed, the EoS parameter of DE prevails to the quintessence region, crossing the phantom-divide has been realized at present, namely KHDE favors the phantom region. However, the most attractive characteristic is that comparing to the flat case, where the phantom regime was not obtained, the incorporation of curvature leads the Universe to the phantom regime. This is an advantage since one expects that only small deviations from standard entropy could be one of the cases \\

It is however normal to develop the present study with the extension of other data from BAO (Baryon Acoustic Oscillation), SNIa, and CMB (Cosmic microwave background) probes in order to restrain the new parameters more accurately. Work along this line is in progress and further results on this topic will be reported in forthcoming researches.

\section*{Acknowledgements}
The authors are thankful to Ms. Gunjan Varshney, GLA University, Mathura, India for her help in removing the glitches of the literature.
\bibliography{ref2}

\begin{thebibliography}{10}

\bibitem{gibbs1960elementary}
J.~W. Gibbs, ``Elementary principles in statistical mechanics: 1902,'' {\em New
  York: Charles Scribner’s Sons}, 1960.

\bibitem{kaniadakis2001non}
G.~Kaniadakis, ``Non-linear kinetics underlying generalized statistics,'' {\em
  Physica A: Statistical mechanics and its applications}, vol.~296, no.~3-4,
  pp.~405--425, 2001.

\bibitem{kaniadakis2002statistical}
G.~Kaniadakis, ``Statistical mechanics in the context of special relativity,''
  {\em Physical review E}, vol.~66, no.~5, p.~056125, 2002.

\bibitem{masi2005step}
M.~Masi, ``A step beyond tsallis and r{\'e}nyi entropies,'' {\em Physics
  Letters A}, vol.~338, no.~3-5, pp.~217--224, 2005.

\bibitem{tsallis2013black}
C.~Tsallis and L.~J. Cirto, ``Black hole thermodynamical entropy,'' {\em The
  European Physical Journal C}, vol.~73, no.~7, p.~2487, 2013.

\bibitem{Moradpour:2016rcy}
H.~Moradpour, ``{Implications, consequences and interpretations of generalized
  entropy in the cosmological setups},'' {\em Int. J. Theor. Phys.}, vol.~55,
  no.~9, pp.~4176--4184, 2016.

\bibitem{Nunes:2015xsa}
R.~C. Nunes, E.~M. Barboza, Jr., E.~M.~C. Abreu, and J.~A. Neto, ``{Probing the
  cosmological viability of non-gaussian statistics},'' {\em JCAP}, vol.~08,
  p.~051, 2016.

\bibitem{Komatsu:2016vof}
N.~Komatsu, ``{Cosmological model from the holographic equipartition law with a
  modified R\'enyi entropy},'' {\em Eur. Phys. J. C}, vol.~77, no.~4, p.~229,
  2017.

\bibitem{Moradpour:2017ycq}
H.~Moradpour, A.~Bonilla, E.~M.~C. Abreu, and J.~A. Neto, ``{Accelerated cosmos
  in a nonextensive setup},'' {\em Phys. Rev. D}, vol.~96, no.~12, p.~123504,
  2017.

\bibitem{Abreu:2018pua}
E.~M.~C. Abreu, J.~A. Neto, A.~C.~R. Mendes, A.~Bonilla, and R.~M. de~Paula,
  ``{Tsallis' entropy, modified Newtonian accelerations and the Tully-Fisher
  relation},'' {\em EPL}, vol.~124, no.~3, p.~30005, 2018.

\bibitem{tavayef2018tsallis}
M.~Tavayef, A.~Sheykhi, K.~Bamba, and H.~Moradpour, ``Tsallis holographic dark
  energy,'' {\em Physics Letters B}, vol.~781, pp.~195--200, 2018.

\bibitem{jahromi2018generalized}
A.~S. Jahromi, S.~Moosavi, H.~Moradpour, J.~M. Gra{\c{c}}a, I.~Lobo, I.~Salako,
  and A.~Jawad, ``Generalized entropy formalism and a new holographic dark
  energy model,'' {\em Physics Letters B}, vol.~780, pp.~21--24, 2018.

\bibitem{Sharma:2020ylh}
U.~K. Sharma, G.~Varshney, and V.~C. Dubey, ``{Barrow agegraphic dark
  energy},'' {\em Int. J. Mod. Phys. D}, vol.~30, no.~03, p.~2150021, 2021.

\bibitem{moradpour2018thermodynamic}
H.~Moradpour, S.~Moosavi, I.~Lobo, J.~M. Gra{\c{c}}a, A.~Jawad, and I.~Salako,
  ``Thermodynamic approach to holographic dark energy and the r{\'e}nyi
  entropy,'' {\em The European Physical Journal C}, vol.~78, no.~10, p.~829,
  2018.

\bibitem{Srivastava:2020cyk}
S.~Srivastava and U.~K. Sharma, ``{Barrow holographic dark energy with Hubble
  horizon as IR cutoff},'' {\em Int. J. Geom. Meth. Mod. Phys.}, vol.~18,
  no.~01, p.~2150014, 2021.

\bibitem{biro2013q}
T.~S. Bir{\'o} and V.~G. Czinner, ``A q-parameter bound for particle spectra
  based on black hole thermodynamics with r{\'e}nyi entropy,'' {\em Physics
  Letters B}, vol.~726, no.~4-5, pp.~861--865, 2013.

\bibitem{ghaffari2019black}
S.~Ghaffari, A.~Ziaie, H.~Moradpour, F.~Asghariyan, F.~Feleppa, and M.~Tavayef,
  ``Black hole thermodynamics in sharma--mittal generalized entropy
  formalism,'' {\em General Relativity and Gravitation}, vol.~51, no.~7,
  pp.~1--11, 2019.

\bibitem{ghaffari2020inflation}
S.~Ghaffari, A.~Ziaie, V.~Bezerra, and H.~Moradpour, ``Inflation in the
  r{\'e}nyi cosmology,'' {\em Modern Physics Letters A}, vol.~35, no.~01,
  p.~1950341, 2020.

\bibitem{Moradpour:2019wpj}
H.~Moradpour, A.~H. Ziaie, S.~Ghaffari, and F.~Feleppa, ``{The generalized and
  extended uncertainty principles and their implications on the Jeans mass},''
  {\em Mon. Not. Roy. Astron. Soc.}, vol.~488, no.~1, pp.~L69--L74, 2019.

\bibitem{Moradpour:2018ima}
H.~Moradpour, A.~Sheykhi, C.~Corda, and I.~G. Salako, ``{Implications of the
  generalized entropy formalisms on the Newtonian gravity and dynamics},'' {\em
  Phys. Lett. B}, vol.~783, pp.~82--85, 2018.

\bibitem{Moradpour:2019yiq}
H.~Moradpour, C.~Corda, A.~H. Ziaie, and S.~Ghaffari, ``{The extended
  uncertainty principle inspires the R\'enyi entropy},'' {\em EPL}, vol.~127,
  no.~6, p.~60006, 2019.

\bibitem{Barrow:2020tzx}
J.~D. Barrow, ``{The Area of a Rough Black Hole},'' {\em Phys. Lett. B},
  vol.~808, p.~135643, 2020.

\bibitem{Srednicki:1993im}
M.~Srednicki, ``{Entropy and area},'' {\em Phys. Rev. Lett.}, vol.~71,
  pp.~666--669, 1993.

\bibitem{Das:2007sj}
S.~Das and S.~Shankaranarayanan, ``{Where are the black hole entropy degrees of
  freedom?},'' {\em Class. Quant. Grav.}, vol.~24, pp.~5299--5306, 2007.

\bibitem{pavon2020degrees}
D.~Pavon, ``On the degrees of freedom of a black hole,'' {\em arXiv preprint
  arXiv:2001.05716}, 2020.

\bibitem{li2004model}
M.~Li, ``A model of holographic dark energy,'' {\em Physics Letters B},
  vol.~603, no.~1-2, pp.~1--5, 2004.

\bibitem{Myung:2007pn}
Y.~S. Myung, ``{Instability of holographic dark energy models},'' {\em Phys.
  Lett. B}, vol.~652, pp.~223--227, 2007.

\bibitem{Hayward:1997jp}
S.~A. Hayward, ``{Unified first law of black hole dynamics and relativistic
  thermodynamics},'' {\em Class. Quant. Grav.}, vol.~15, pp.~3147--3162, 1998.

\bibitem{Hayward:1998ee}
S.~A. Hayward, S.~Mukohyama, and M.~C. Ashworth, ``{Dynamic black hole
  entropy},'' {\em Phys. Lett. A}, vol.~256, pp.~347--350, 1999.

\bibitem{bak2000cosmic}
D.~Bak and S.-J. Rey, ``Cosmic holography+,'' {\em Classical and Quantum
  Gravity}, vol.~17, no.~15, p.~L83, 2000.

\bibitem{cai2005first}
R.-G. Cai and S.~P. Kim, ``First law of thermodynamics and friedmann equations
  of friedmann-robertson-walker universe,'' {\em Journal of High Energy
  Physics}, vol.~2005, no.~02, p.~050, 2005.

\bibitem{akbar2007thermodynamic}
M.~Akbar and R.-G. Cai, ``Thermodynamic behavior of the friedmann equation at
  the apparent horizon of the frw universe,'' {\em Physical Review D}, vol.~75,
  no.~8, p.~084003, 2007.

\bibitem{cai2009hawking}
R.-G. Cai, L.-M. Cao, and Y.-P. Hu, ``Hawking radiation of an apparent horizon
  in a frw universe,'' {\em Classical and Quantum Gravity}, vol.~26, no.~15,
  p.~155018, 2009.

\bibitem{huang2019stability}
Q.~Huang, H.~Huang, J.~Chen, L.~Zhang, and F.~Tu, ``Stability analysis of a
  tsallis holographic dark energy model,'' {\em Classical and Quantum Gravity},
  vol.~36, no.~17, p.~175001, 2019.

\bibitem{universe7030067}
S.~H. Shekh, P.~H. R.~S. Moraes, and P.~K. Sahoo, ``Physical acceptability of
  the renyi, tsallis and sharma-mittal holographic dark energy models in the
  f(t,b) gravity under hubble’s cutoff,'' {\em Universe}, vol.~7, no.~3,
  2021.

\bibitem{ZUBAIR2021153}
M.~Zubair and L.~R. Durrani, ``Exploring tsallis holographic dark energy
  scenario in f(r,t) gravity,'' {\em Chinese Journal of Physics}, vol.~69,
  pp.~153--171, 2021.

\bibitem{Bhattacharjee_2021}
S.~Bhattacharjee, ``Growth rate and configurational entropy in tsallis
  holographic dark energy,'' {\em The European Physical Journal C}, vol.~81,
  Mar 2021.

\bibitem{nauenberg2003critique}
M.~Nauenberg, ``Critique of q-entropy for thermal statistics,'' {\em Physical
  Review E}, vol.~67, no.~3, p.~036114, 2003.

\bibitem{moyano2003zeroth}
L.~G. Moyano, F.~Baldovin, and C.~Tsallis, ``Zeroth principle of thermodynamics
  in aging quasistationary states,'' {\em arXiv preprint cond-mat/0305091},
  2003.

\bibitem{tsallis2004comment}
C.~Tsallis, ``Comment on “critique of q-entropy for thermal statistics”,''
  {\em Physical Review E}, vol.~69, no.~3, p.~038101, 2004.

\bibitem{nauenberg2004reply}
M.~Nauenberg, ``Reply to “comment on ‘critique of q-entropy for thermal
  statistics’”,'' {\em Physical Review E}, vol.~69, no.~3, p.~038102, 2004.

\bibitem{li2005different}
W.~Li, Q.~A. Wang, L.~Nivanen, and A.~Le~M{\'e}haut{\'e}, ``On different
  q-systems in nonextensive thermostatistics,'' {\em The European Physical
  Journal B-Condensed Matter and Complex Systems}, vol.~48, no.~1, pp.~95--100,
  2005.

\bibitem{abe2006temperature}
S.~Abe, ``Temperature of nonextensive systems: Tsallis entropy as clausius
  entropy,'' {\em Physica A: Statistical Mechanics and its Applications},
  vol.~368, no.~2, pp.~430--434, 2006.

\bibitem{biro2011zeroth}
T.~Bir{\'o} and P.~V{\'a}n, ``Zeroth law compatibility of nonadditive
  thermodynamics,'' {\em Physical Review E}, vol.~83, no.~6, p.~061147, 2011.

\bibitem{biro2011publisher}
T.~Bir{\'o} and P.~V{\'a}n, ``Publisher’s note: Zeroth law compatibility of
  nonadditive thermodynamics [phys. rev. e 83, 061147 (2011)],'' {\em Physical
  Review E}, vol.~84, no.~1, p.~019902, 2011.

\bibitem{abreu2018cosmological}
E.~M. Abreu, J.~A. Neto, A.~C. Mendes, A.~Bonilla, and R.~M. de~Paula,
  ``Cosmological considerations in kaniadakis statistics,'' {\em EPL
  (Europhysics Letters)}, vol.~124, no.~3, p.~30003, 2018.

\bibitem{moradpour2020generalized}
H.~Moradpour, A.~Ziaie, and M.~K. Zangeneh, ``Generalized entropies and
  corresponding holographic dark energy models,'' {\em The European Physical
  Journal C}, vol.~80, no.~8, pp.~1--7, 2020.

\bibitem{Jawad:2021xsr}
A.~Jawad and A.~M. Sultan, ``{Cosmic Consequences of Kaniadakis and Generalized
  Tsallis Holographic Dark Energy Models in the Fractal Universe},'' {\em Adv.
  High Energy Phys.}, vol.~2021, p.~5519028, 2021.

\bibitem{Vagnozzi:2020dfn}
S.~Vagnozzi, A.~Loeb, and M.~Moresco, ``{Eppur \`e piatto? The Cosmic
  Chronometers Take on Spatial Curvature and Cosmic Concordance},'' {\em
  Astrophys. J.}, vol.~908, no.~1, p.~84, 2021.

\bibitem{Aghanim:2018eyx}
N.~Aghanim {\em et~al.}, ``{Planck 2018 results. VI. Cosmological
  parameters},'' {\em Astron. Astrophys.}, vol.~641, p.~A6, 2020.

\bibitem{Vagnozzi:2020zrh}
S.~Vagnozzi, E.~Di~Valentino, S.~Gariazzo, A.~Melchiorri, O.~Mena, and J.~Silk,
  ``{Listening to the BOSS: the galaxy power spectrum take on spatial curvature
  and cosmic concordance},'' 10 2020.

\bibitem{DiValentino:2020hov}
E.~Di~Valentino, A.~Melchiorri, and J.~Silk, ``{Investigating Cosmic
  Discordance},'' {\em Astrophys. J. Lett.}, vol.~908, no.~1, p.~L9, 2021.

\bibitem{sheykhi2018modified}
A.~Sheykhi, ``Modified friedmann equations from tsallis entropy,'' {\em Physics
  Letters B}, vol.~785, pp.~118--126, 2018.

\bibitem{hayward1999dynamic}
S.~A. Hayward, S.~Mukohyama, and M.~Ashworth, ``Dynamic black-hole entropy,''
  {\em Physics Letters A}, vol.~256, no.~5-6, pp.~347--350, 1999.

\bibitem{hayward1998unified}
S.~A. Hayward, ``Unified first law of black-hole dynamics and relativistic
  thermodynamics,'' {\em Classical and Quantum Gravity}, vol.~15, no.~10,
  p.~3147, 1998.

\bibitem{moresco20166}
M.~Moresco, L.~Pozzetti, A.~Cimatti, R.~Jimenez, C.~Maraston, L.~Verde,
  D.~Thomas, A.~Citro, R.~Tojeiro, and D.~Wilkinson, ``A 6\% measurement of the
  hubble parameter at z~ 0.45: direct evidence of the epoch of cosmic
  re-acceleration,'' {\em Journal of Cosmology and Astroparticle Physics},
  vol.~2016, no.~05, p.~014, 2016.

\bibitem{suzuki2012hubble}
N.~Suzuki, D.~Rubin, C.~Lidman, G.~Aldering, R.~Amanullah, K.~Barbary,
  L.~Barrientos, J.~Botyanszki, M.~Brodwin, N.~Connolly, {\em et~al.}, ``The
  hubble space telescope cluster supernova survey. v. improving the dark-energy
  constraints above z> 1 and building an early-type-hosted supernova sample,''
  {\em The Astrophysical Journal}, vol.~746, no.~1, p.~85, 2012.

\bibitem{liddle2000cosmological}
A.~R. Liddle and D.~H. Lyth, {\em Cosmological inflation and large-scale
  structure}.
\newblock Cambridge university press, 2000.

\bibitem{copeland2006m}
E.~Copeland, ``M sami and s tsujikawa int. j. mod,'' {\em Phys. D}, vol.~15,
  p.~1753, 2006.

\bibitem{calabrese2011future}
E.~Calabrese, R.~de~Putter, D.~Huterer, E.~V. Linder, and A.~Melchiorri,
  ``Future cmb constraints on early, cold, or stressed dark energy,'' {\em
  Physical Review D}, vol.~83, no.~2, p.~023011, 2011.

\bibitem{vagnozzi2020we}
S.~Vagnozzi, L.~Visinelli, O.~Mena, and D.~F. Mota, ``Do we have any hope of
  detecting scattering between dark energy and baryons through cosmology?,''
  {\em Monthly Notices of the Royal Astronomical Society}, vol.~493, no.~1,
  pp.~1139--1152, 2020.

\end{thebibliography}
\bibliographystyle{ieeetr}

\end{document}